\documentclass[aps,prl,superscriptaddress]{revtex4}
\usepackage{amsmath}
\usepackage{amssymb}
\usepackage{graphicx}
\usepackage{dcolumn}
\usepackage{bm,CJK}

\begin{document}

\title{Dynamical Universal Behavior in Quantum Chaotic Systems}
\author{Hongwei Xiong}
\affiliation{State Key Laboratory of Magnetic Resonance and Atomic and Molecular Physics,
Wuhan Institute of Physics and Mathematics, Chinese Academy of Sciences,
Wuhan 430071, China}
\author{Biao Wu}
\email{wubiao.phys@gmail.com}
\affiliation{International Center for Quantum Materials, Peking University, Beijing
100871, China}
\affiliation{Institute of Physics, Chinese Academy of Sciences, Beijing
100190, China}

\date{June 21st, 2010}

\begin{abstract}
We discover numerically that a moving wave packet in a quantum chaotic
billiard will always evolve into a quantum state, whose density probability
distribution is exponential. This exponential distribution is found to be
universal for quantum chaotic systems with rigorous
proof. In contrast, for the corresponding classical system, the distribution is Gaussian.
We find that the quantum exponential distribution can smoothly change to
the classical Gaussian  distribution with coarse graining.
\end{abstract}

\maketitle



\bigskip

The study of quantum chaos intends to elucidate the intriguing
correspondence between quantum mechanics and classical mechanics by focusing
on the effects of chaos of a classical system on its quantum counterpart
\cite{GutzwillerBook,StockmannBook,NatureJensen,Steck2001Science,%
Chaudhury2009Nature}.
Intensive efforts
have found certain universal characteristics of quantum chaotic systems, for
example, the Wigner distribution of energy level spacing \cite{StockmannBook} and the
\textquotedblleft scarring" of eigenstates along the classical periodic
orbits\cite{Heller1984PRL,Wilkinson1996Nature}.  By following a pioneering
work  \cite{Tomsovic1993PRE}, we  study the dynamical evolution of a
moving wave packet inside a ripple quantum billiard \cite{Li2002PRE}.
We find that the wave packet evolution always leads to an
\textquotedblleft equilibrium" state, where the density probability is
exponentially distributed. With rigorous proof, we show that this
exponential distribution is universal for all quantum chaotic systems. This
exponential law is in stark contrast with the evolution of a
\textquotedblleft cloud" of classical particles, which always leads to a Gaussian
distribution. This universal behavior is observable with Bose-Einstein
condensates (BECs) \cite{Dalfovo1999RMP,Chin2009Nature} and other quantum
systems. Furthermore, we demonstrate that the quantum-classical correspondence
between the exponential distribution  and  the Gaussian distribution can be
achieved with coarse graining.

\begin{figure}[tbp]
\includegraphics[width=3.5cm,angle=0]{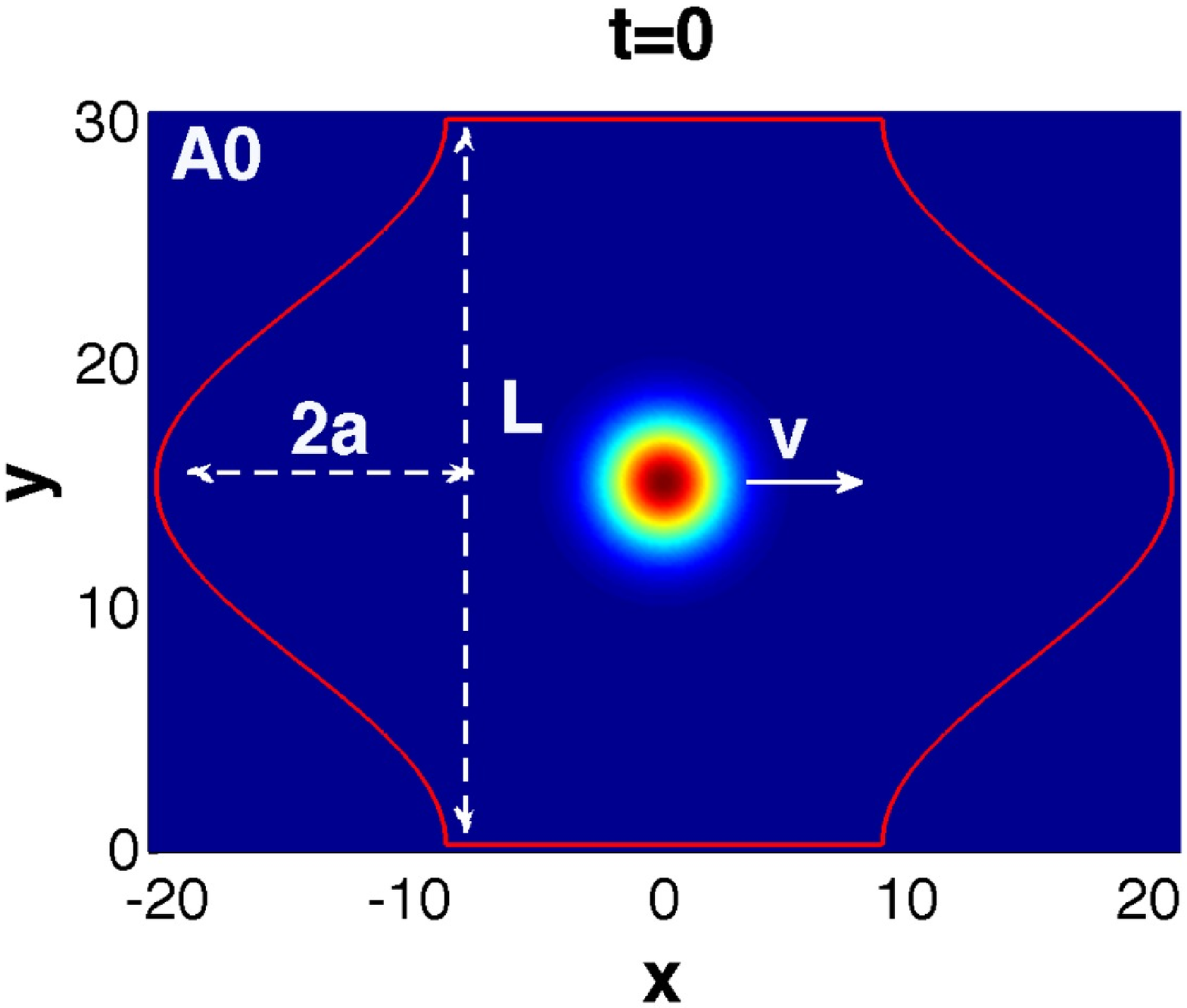} %
\includegraphics[width=3.5cm,angle=0]{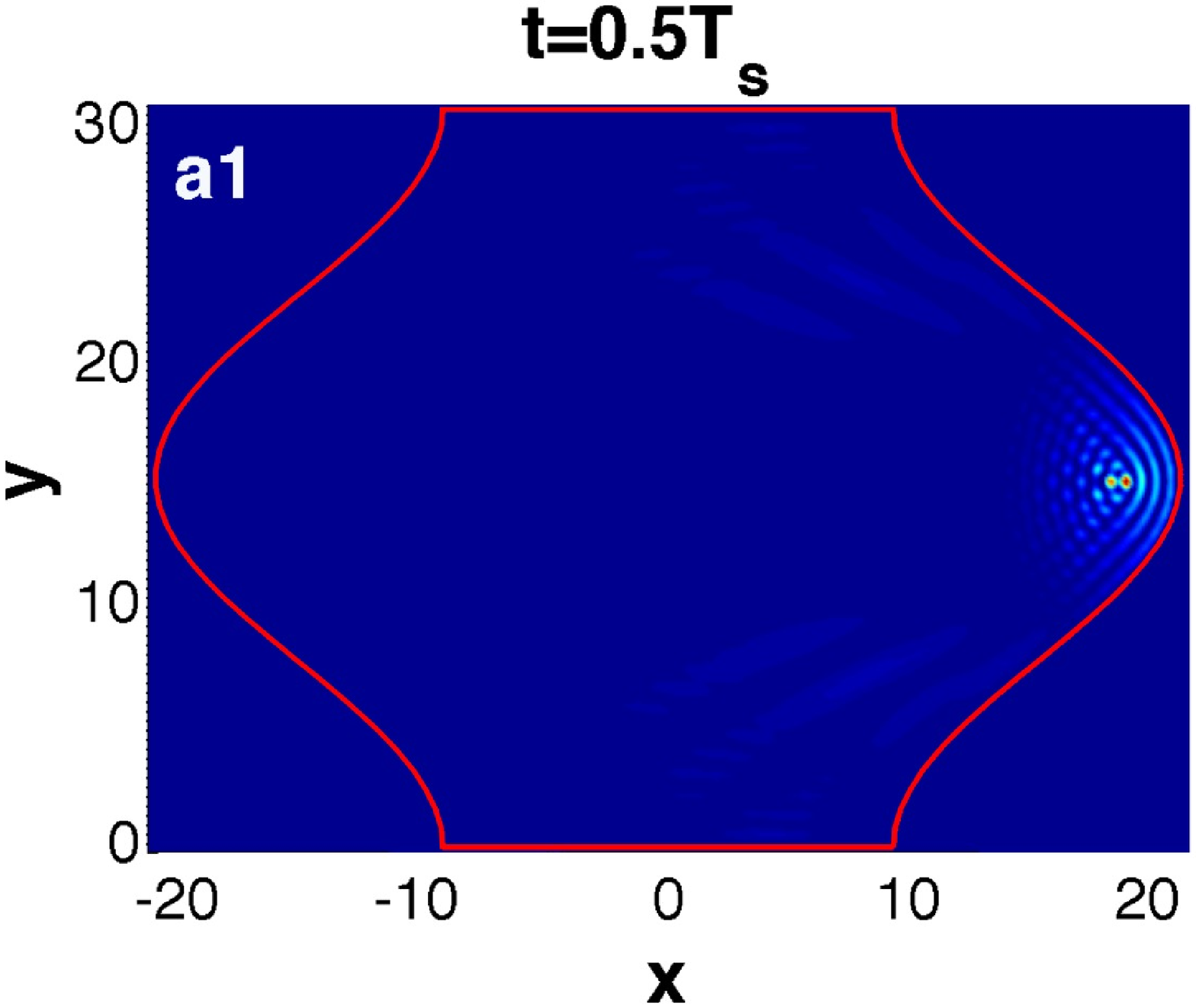} %
\includegraphics[width=3.5cm,angle=0]{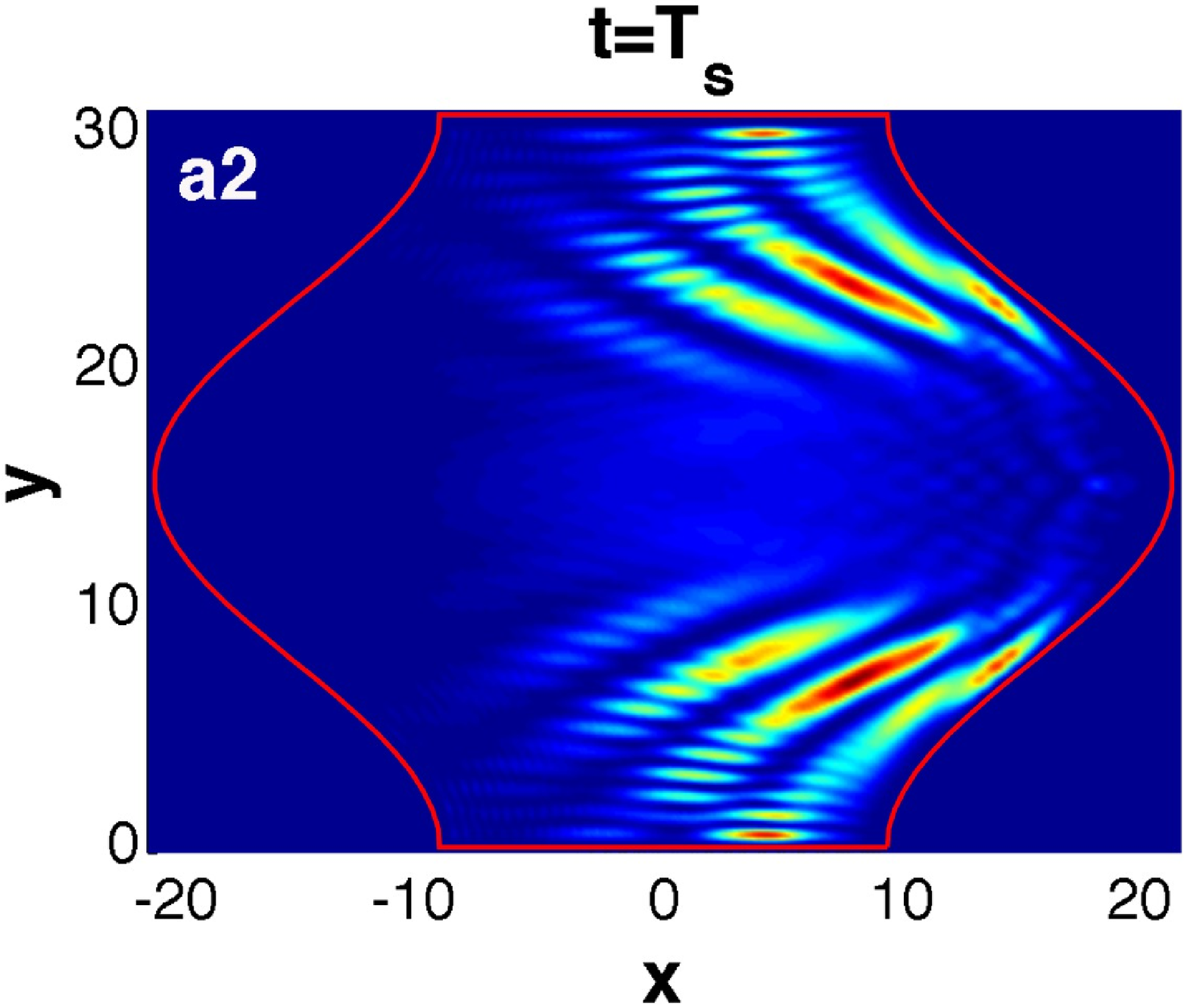} %
\includegraphics[width=3.5cm,angle=0]{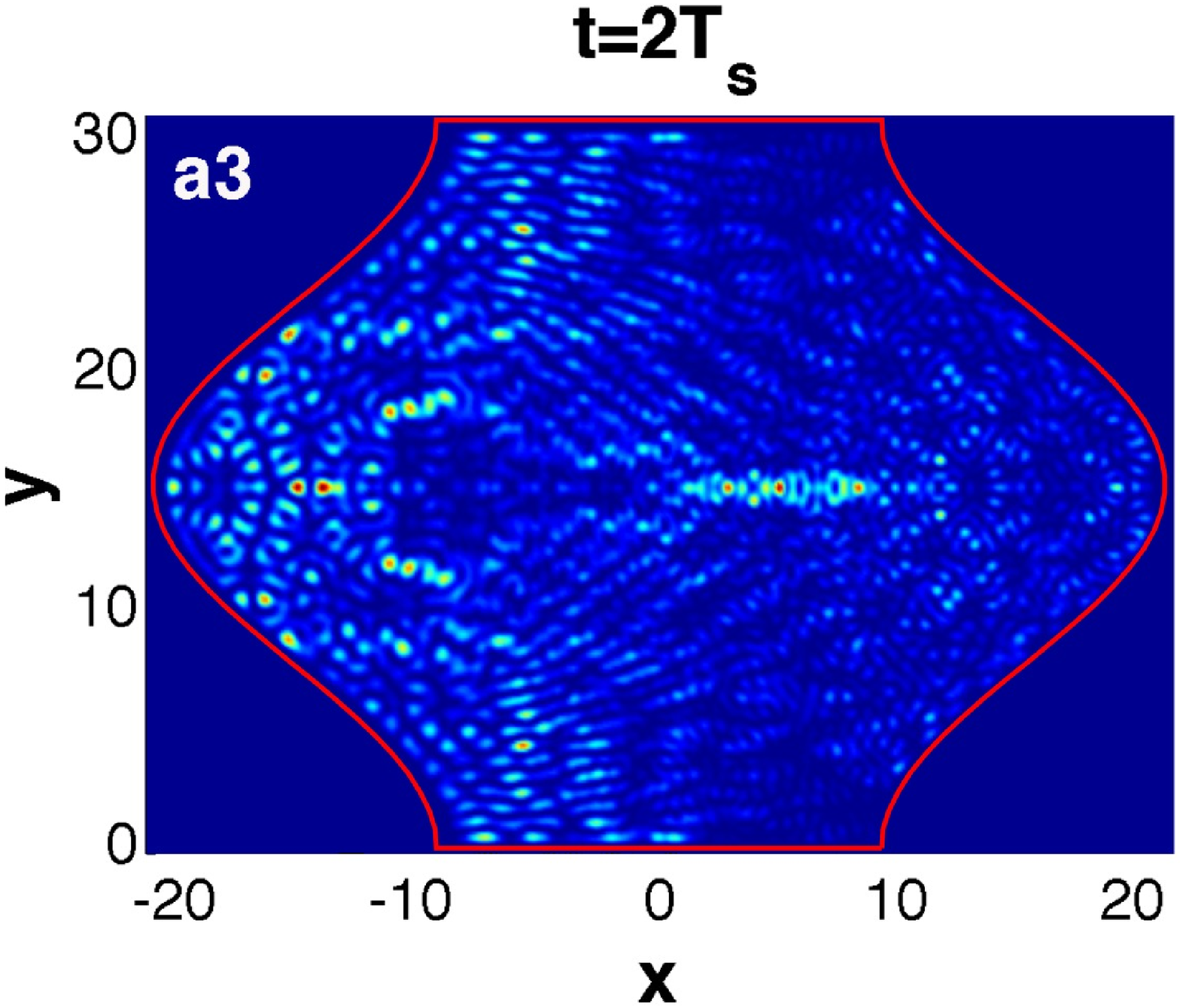} %
\includegraphics[width=3.5cm,angle=0]{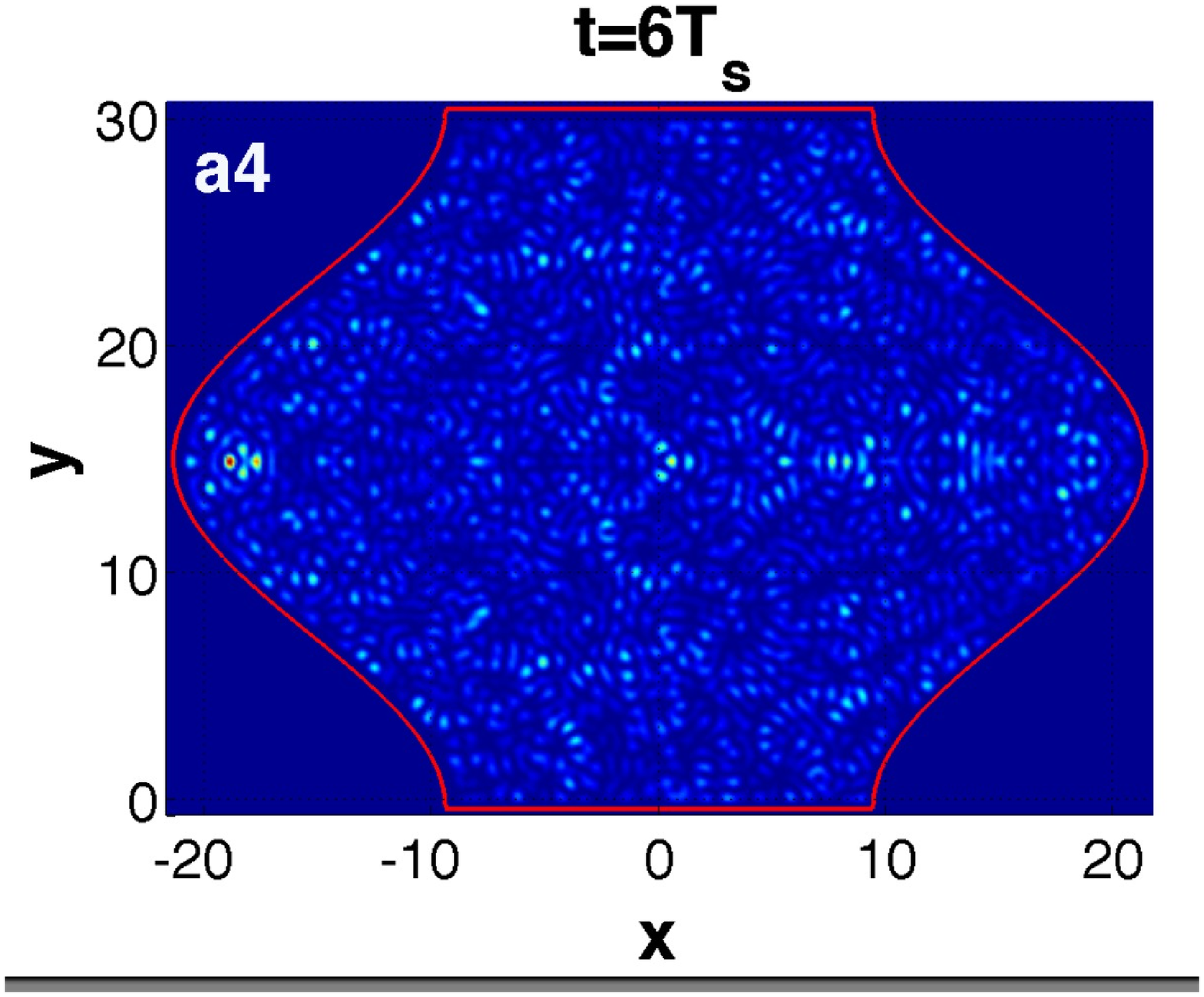}\newline
\includegraphics[width=3.5cm,angle=0]{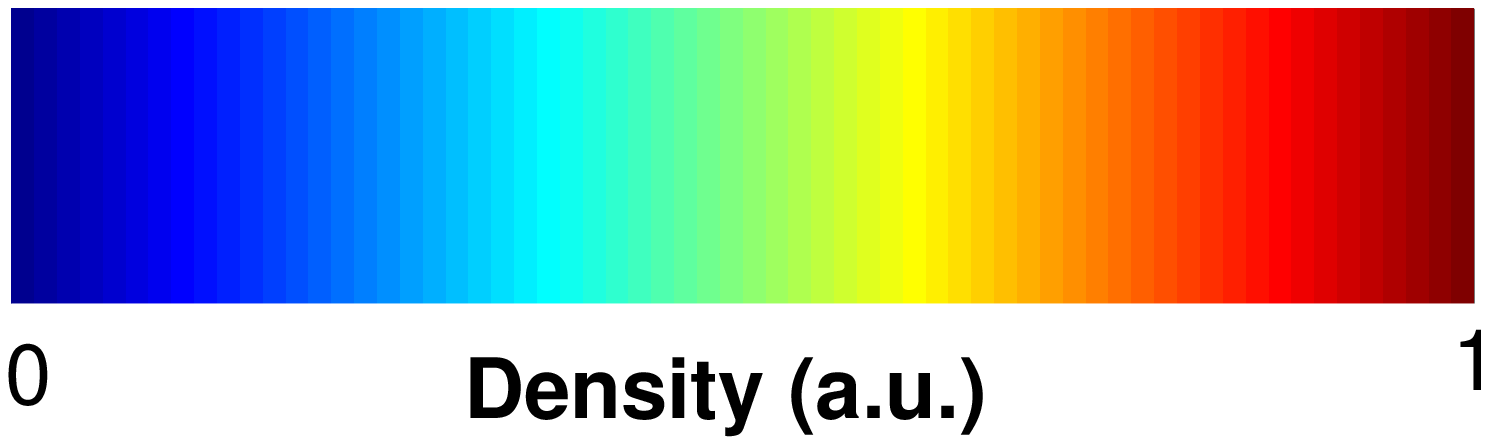} %
\includegraphics[width=3.5cm,angle=0]{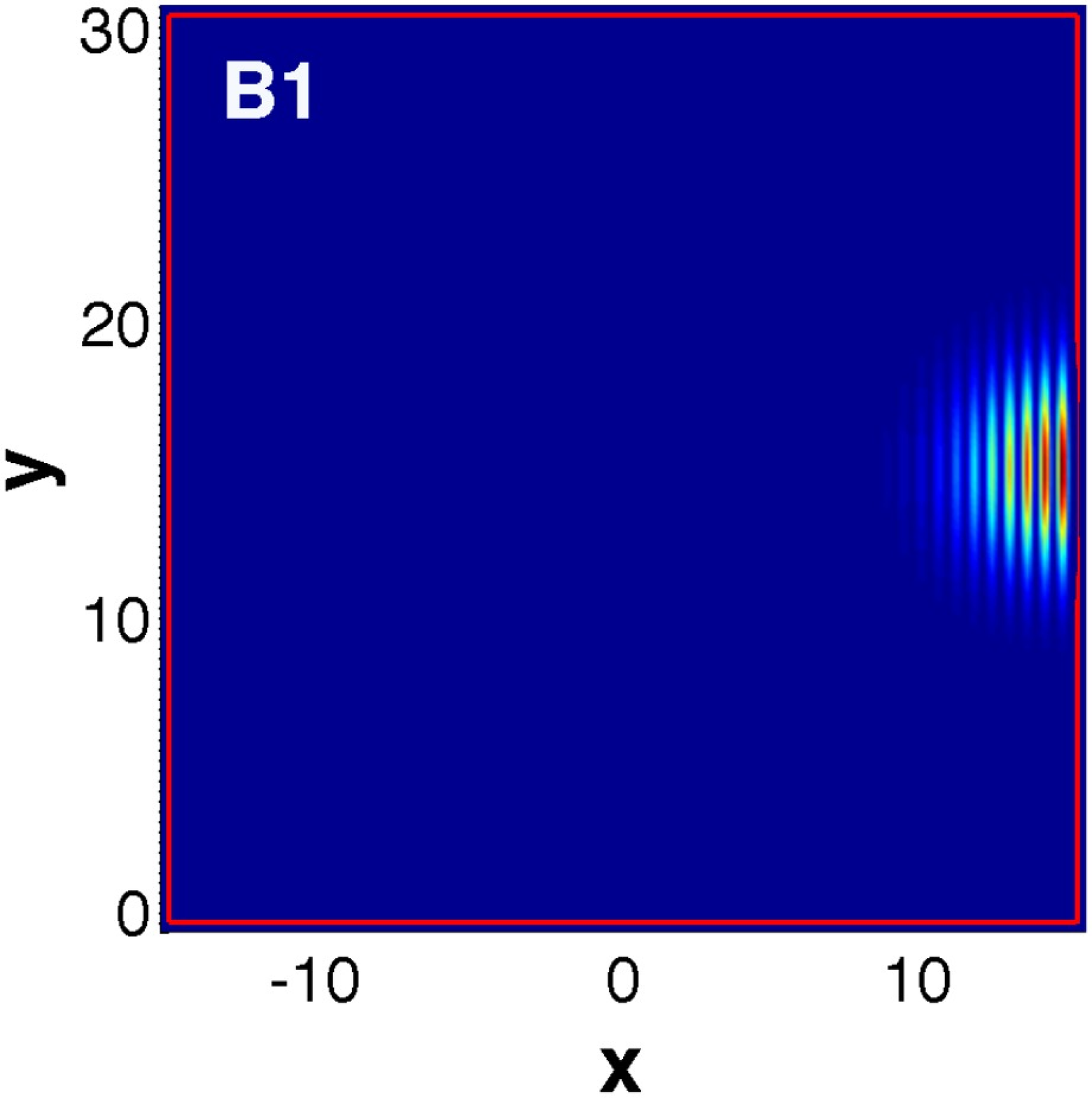} %
\includegraphics[width=3.5cm,angle=0]{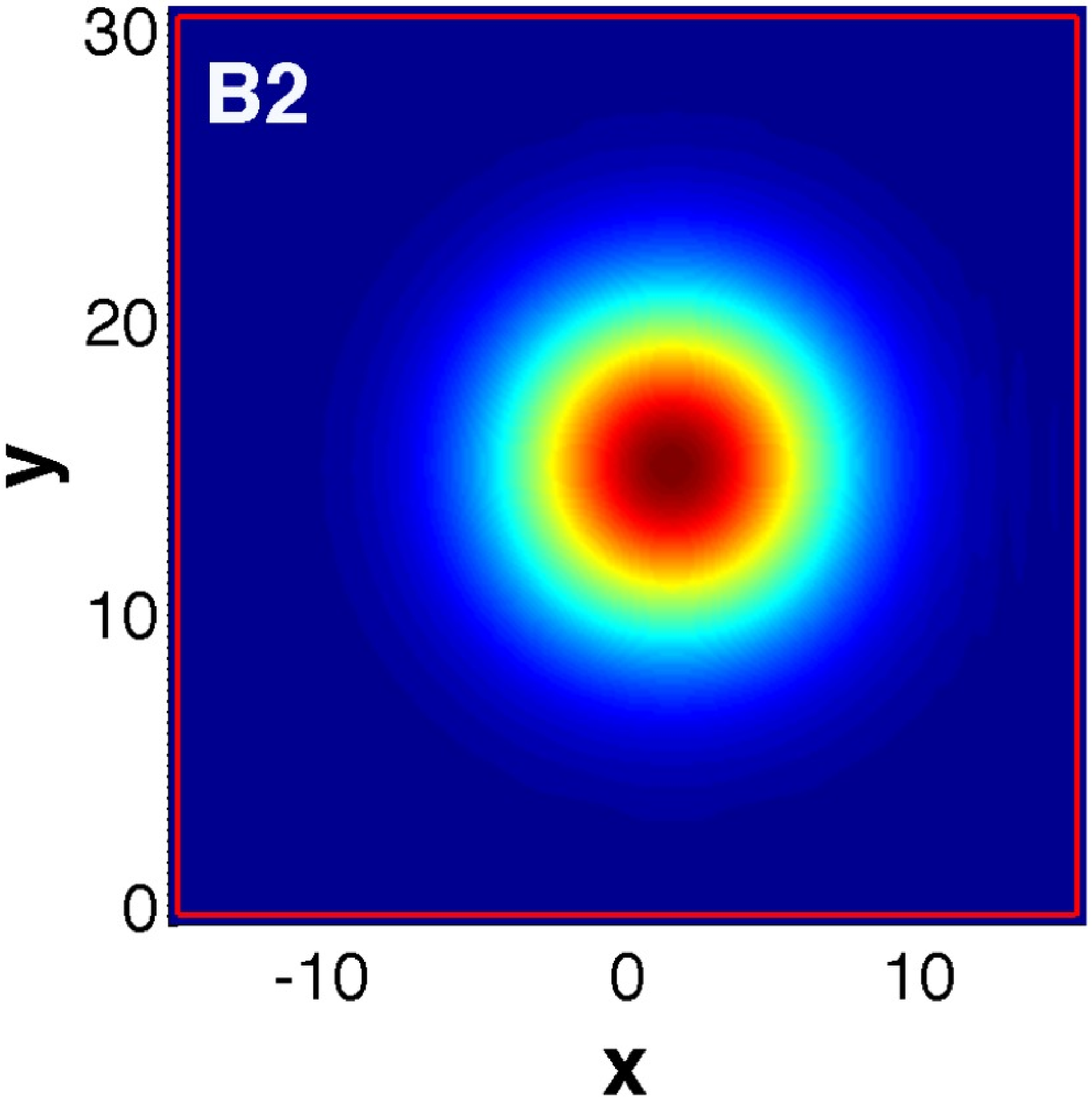} %
\includegraphics[width=3.5cm,angle=0]{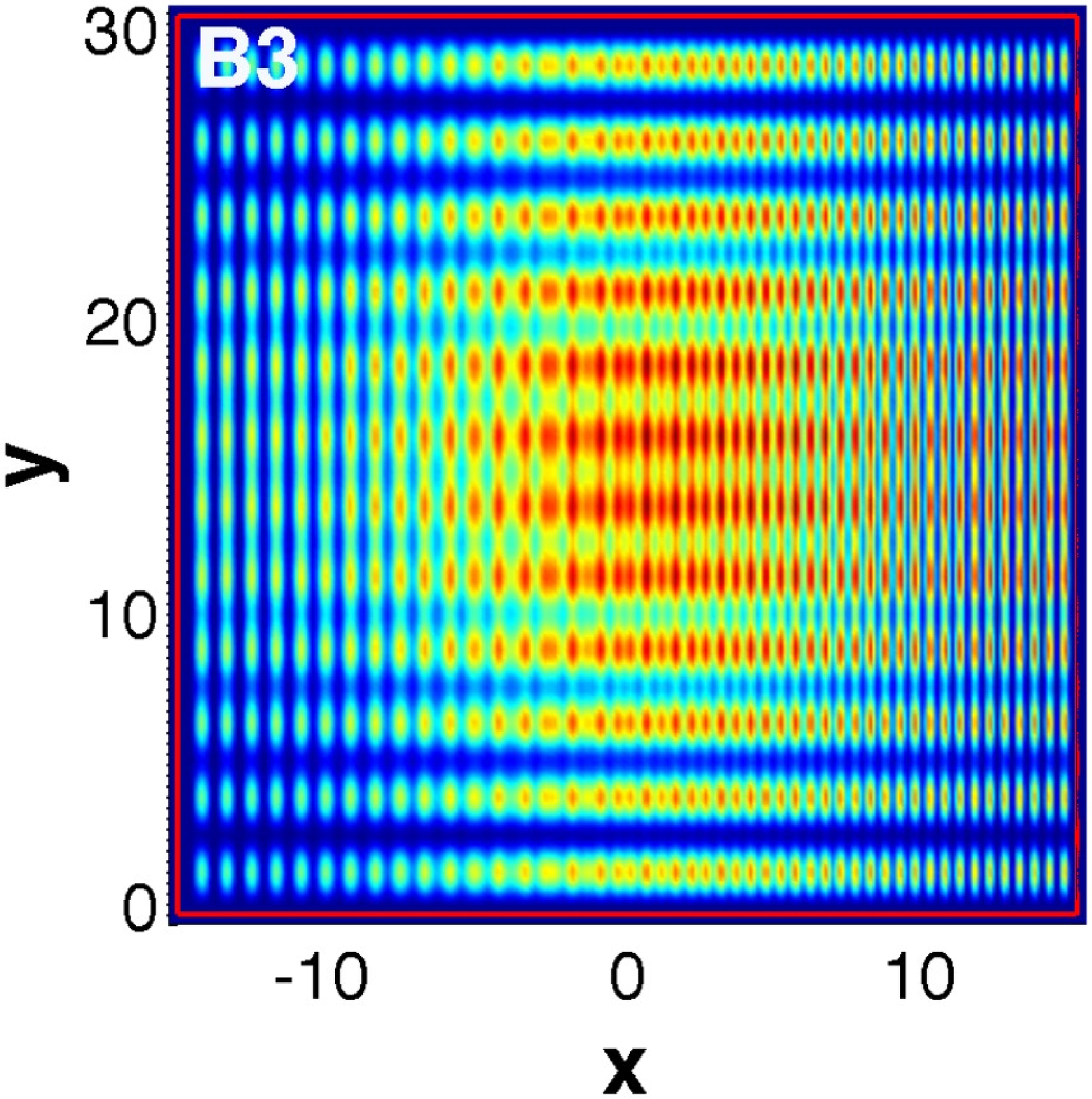} %
\includegraphics[width=3.5cm,angle=0]{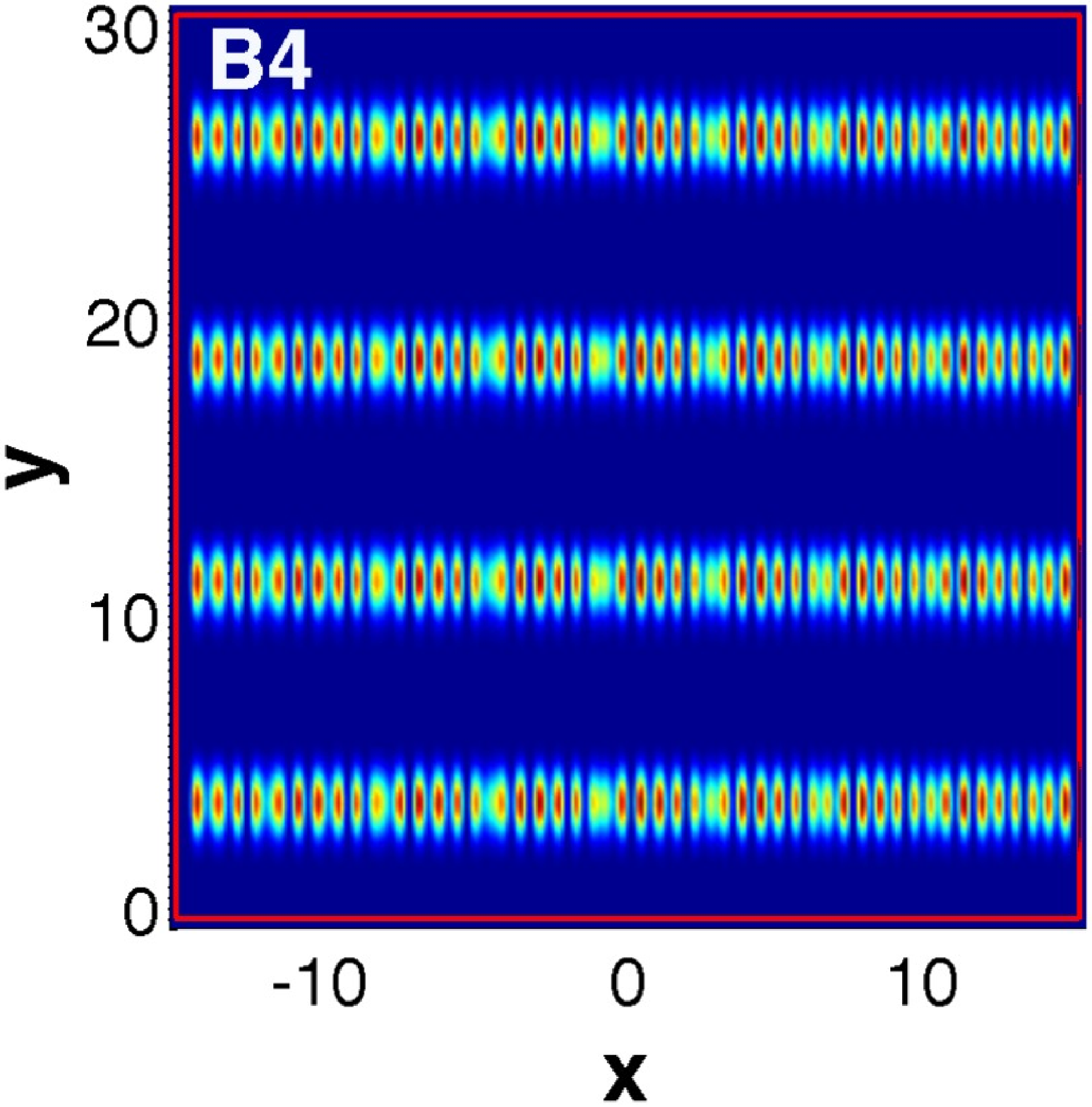}\\
\vspace{5pt}
\includegraphics[width=4.5cm,angle=0]{saturation} \hspace{1cm}%
\includegraphics[width=4.5cm,angle=0]{ripple_dis}%
\caption{Dynamical evolution of a moving wave packet in a ripple
billiard and a square billiard. (\textbf{A0-A4}) The densities of the wave packet in the
ripple billiard at different times $t$. (\textbf{B1}-\textbf{B4}) The densities of the wave packet 
in the square billiard at the same set of times except at $t=0$. 
(\textbf{C}) The time evolution of the probability for the density being above the
average. The red-dashed line is for $e^{-1}$ with $e$ being the Euler's
number. (\textbf{D}) The density probability distribution functions at $%
t=12T_{s}$. The red solid line is for $\exp (-n_{0})$. The inset shows the
semilog plot of the same sets of data.
For the ripple billiard, the
parameters are $L=30$, $b=15$, and $a=6$. The initial speed of the wave
packet is $v_0=10$. $t_{0}$ is in unit of $T_{s}=2(a+b)/v_0$ and $n_{0}$
is in unit of the averaged density $n_{s}$. $\hbar =2m=1$ are
adopted in the compuation. $\protect\delta n=0.02$ is used to obtain the
results in (\textbf{D}).}
\label{fig1}
\end{figure}

We consider the following Schr\"{o}dinger equation with the units $\hbar
=2m=1$%
\begin{equation}
i\frac{\partial \Psi }{\partial t}=\left( -\frac{\partial ^{2}}{\partial
x^{2}}-\frac{\partial ^{2}}{\partial y^{2}}+V_{b}\right) \Psi ,
\label{NSequation}
\end{equation}%
where $V_{b}$ represents the hard-wall potential for a ripple billiard
(see Fig. \ref{fig1}A0). The ripple billiards have the advantage of
having an exact Hamiltonian matrix in terms of elementary functions \cite%
{Li2002PRE}. The left and right walls are described by functions $%
\mp(b-a\cos \left( 2\pi y/L\right) )$, respectively. The initial wave
function is a moving Gaussian wave packet given by $\Psi \left(
x,y,t=0\right) =\Psi _{G}\left( x,y\right) e^{iv_0x/2}$, where $\Psi
_{G}\left( x,y\right) $ is the ground-state wave function in a harmonic trap
$V_{h}=\left( \omega _{x}^{2}x^{2}+\omega _{y}^{2}y^{2}\right) /4$.\newline

The time evolution of the wave packet density $n\left( x,y,t\right)
=\left\vert \Psi \left( x,y,t\right) \right\vert ^{2}$ is shown in Figs.\ref%
{fig1}A1-A4 for the parameters $\omega _{x}=\omega _{y}=2$, $v_0=10$, $%
L=30$, $b=15$ and $a=6$. The time is in the unit of $T_{s}=2\left(a+b\right)
/v_0$, which is the longest time for a classical particle making a
round-trip along the $x$ direction inside the billiard. As clearly seen in
the figure, the wave packet moves and expands, and eventually gets reflected multiple times
by the curved hard walls. As a result, the wave packet gets smeared out in
the billiard. After about $t=6T_s$, the density of the wave packet reaches kind of
an ``equilibrium" state, where the overall features of the wave packet no
longer changes. For comparison, the evolution of the same wave packet in a
square billiard ($a=0$) is shown in Figs.\ref{fig1}B1-B4. The difference is
obvious: in the square billiard, the wave packet always has certain
patterns, looking very regular, and it never settles into any fixed pattern.
\newline

With the inspiration to describe quantitatively the contrasting images
between Figs.\ref{fig1}A3-A4 and Figs.\ref{fig1}B3-B4, we introduce the
density probability distribution function $P\left(n,t\right) $, which is
defined as the probability of the wave packet density being between $%
n-\delta n/2$ and $n+\delta n/2$. Its mathematical expression is
\begin{equation}
P\left( n,t\right) =\frac{S\left( n-\delta n/2,t\right) -S\left( n+\delta
n/2,t\right) }{\delta n\cdot S_{\mathrm{total}}}\,,  \label{defin}
\end{equation}%
%
%
%
%
where $S_{\mathrm{total}}$ is the area of the whole billiard and $S\left(
n,t\right) $ is the area of the regions, where the density is larger than $n$%
. This density probability distribution is normalized by definition. As the
wave packet in our calculation is normalized, the averaged density is $%
n_{s}=1/S_{\mathrm{total}}$. With the averaged density, we define a
dimensionless density as $n_{0}=n/n_{s}$. The dimensionless probability
distribution function is then $P_{0}\left( n_{0},t_{0}\right) =n_s P\left(
n,t\right)$ with $t_{0}=t/T_{s}$.\\

Figure \ref{fig1}C  shows the time evolution of $s_{a}(t)=S\left(
n_{s},t\right) /S_{\mathrm{total}}$, the probability of having density
above the average. For the square billiard ($a=0)$, $s_{a}(t)$ exhibits
large-amplitude oscillations, reflecting the ever-changing regular patterns
in the density. Very differently and also strikingly, $s_{a}$ quickly
reaches a plateau, fluctuating slightly around a fixed value, which is
around $e^{-1}$ with $e\approx 2.718$ being the Euler's number. This signals
that the wave packet finally evolves into an \textquotedblleft equilibrium"
state, where the overall feature of the wave function no longer changes. 
The distribution function $P_{0}\left( n_{0},t_{0}\right) $ also has similar behavior: for
the square billiard, $P_{0}$ always changes with time; in contrast, for the
ripple billiard, $P_{0}$ settles  into an \textquotedblleft
equilibrium" function. As shown in Fig. \ref{fig1}D, we find by numerical
fitting that the \textquotedblleft equilibrium" function is exponential,
\begin{equation}
P_{0}^{\mathrm{eq}}\left( n_{0}\right) =e^{-n_{0}}.  \label{univ}
\end{equation}%
This simple exponential law is also found numerically for the the widely
studied stadium billiard \cite{Heller1984PRL,Tomsovic1993PRE}. The evolution
of a wave packet in Fig.\ref{fig1} can also be regarded as an evolution of a
BEC without interaction. In fact, all the parameters chosen for Fig. \ref%
{fig1} come from relevant BEC experiments \cite{Dalfovo1999RMP}. We find
that  for sufficiently large interaction, the density distribution of a BEC
in the square billiard can also reach the above exponential distribution
after long-time evolution. These numerical calculations indicate that the
exponential law in Eq.(\ref{univ}) is likely universal and may apply in any
quantum chaotic systems and in many different settings.  To support this,
we provide two more sets of numerical results.  \newline

\begin{figure}[!htp]
\begin{minipage}{6.0cm}
\includegraphics[width=6.0cm,angle=0,bb=58 218 534 623]{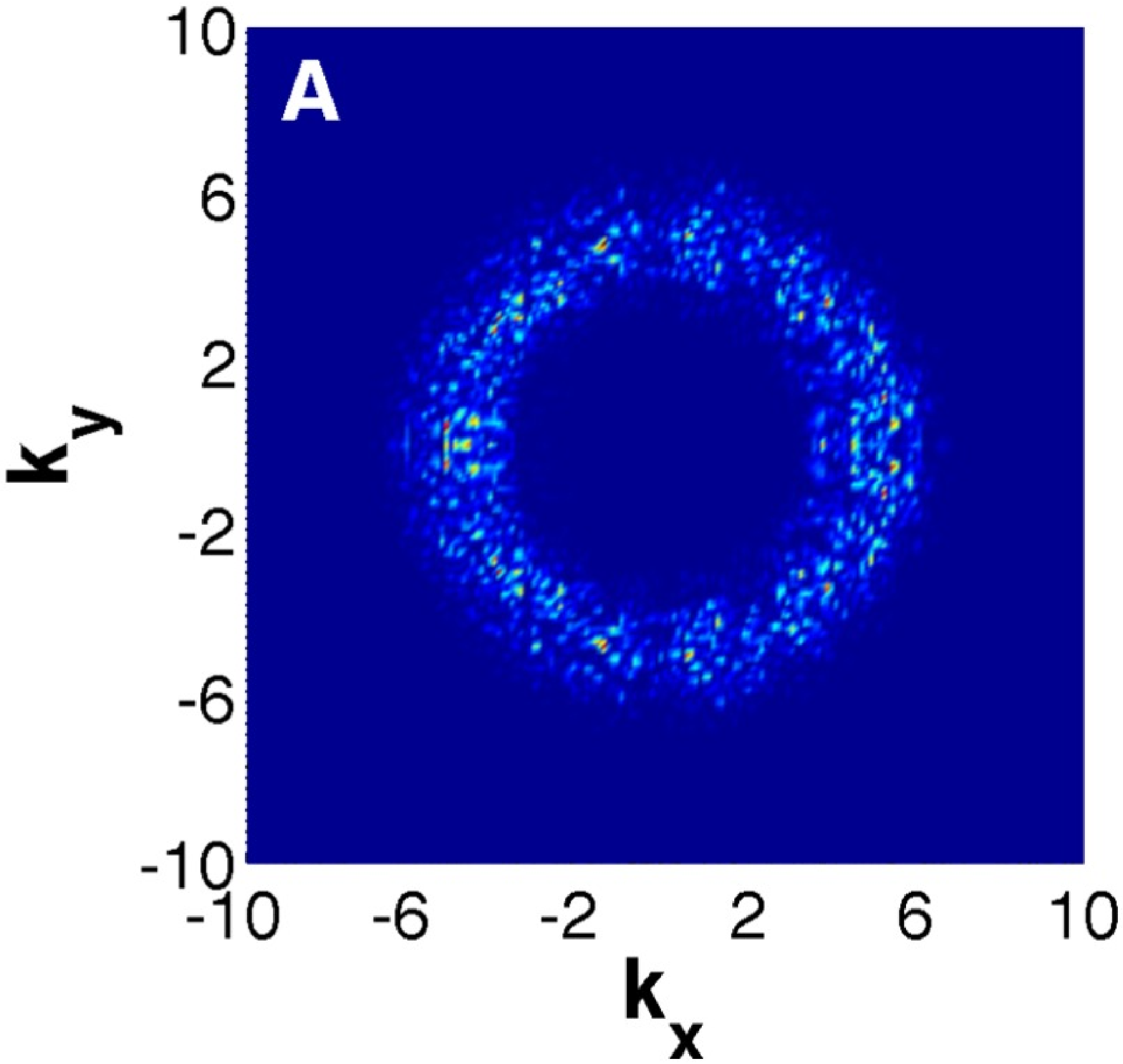} %
\end{minipage}
\begin{minipage}{6.0cm}
\includegraphics[width=4.3cm,angle=0]{momt_dist6T} %
\end{minipage}
\\
\begin{minipage}{6.0cm}
\includegraphics[width=5.3cm,angle=0]{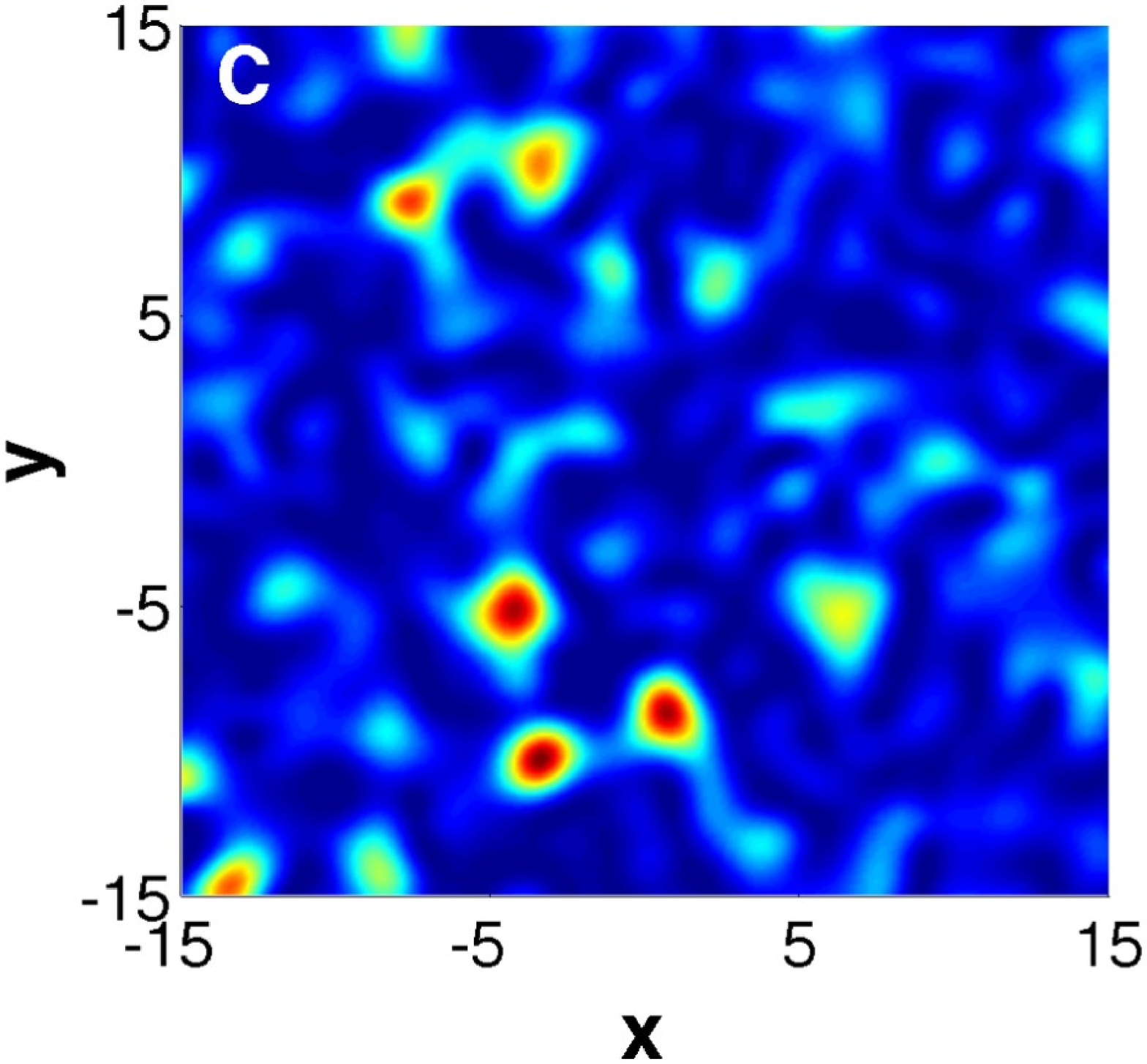}
\end{minipage}
\begin{minipage}{6.0cm}
\includegraphics[width=4.3cm,angle=0]{cgrain}
\end{minipage}
\caption{Two more examples of exponential distributions. 
({\bf A}) Density distribution in the momentum space for the quantum state shown in Fig.\ref{fig1}A4.
({\bf B}) Probability distribution for the momentum distribution in ({\bf A}).
In obtaining ({\bf B}), the momentum space is divided into concentric rings with equal
spacing $\delta k=\pi/30$ and each ring is then divided into equal pieces so that each piece
has the area of $\sim\delta k\times \delta k$.  The density at each unit piece is renormalized with the averaged
density of the ring where the unit piece belongs. The large fluctuations in ({\bf B}) are caused
by the small number of sample points. ({\bf C})  The density of the randomly superposed plane waves.
It is computed according to Eq.(\protect\ref{randomwavefunction})
with 10,000 plane waves for an area of 400$\times$400.
 ({\textbf D}) Density probability distributions  for the random quantum state in ({\bf C}).
The distribution functions are obtained by
dividing this whole area equally into different square
cells, $0.1\times 0.1$(black line), $2\times 2$(green line), and $8\times 8$%
(red line). 
}
\label{fig2}
\end{figure}

In Fig.\ref{fig2}A, we have plotted the momentum
distribution of the wave function shown in Fig.\ref{fig1}A4.  The momentum
distribution has a ring-shaped structure with a spotty look.  To understand
this structure,  let us consider the time evolution of a ``cloud" of non-interacting
identical classical particles, whose initial velocity distributions are identical
to the wave packet in Fig.\ref{fig1}A0, that is,
$G\left( \overrightarrow{v},t=0\right) \sim e^{-\left\vert
\overrightarrow{v}-\overrightarrow{v}_{0}\right\vert ^{2}/2\sigma _{v}^{2}}$.
For a chaotic billiard with hard-wall boundary (e.g., the
ripple billiard), after long-time evolution, the
velocity distribution will become $G\left( \overrightarrow{v},t\gg
T_{s}\right) \sim e^{-\left( \left\vert \overrightarrow{v}\right\vert
^{2}-v_0 ^{2}\right) /2\sigma
_{v}^{2}}$, which is exactly a ring-shaped distribution with a maximum
at $v=v_0$. However, $G\left( \overrightarrow{v},t\gg T_{s}\right)$
is very smooth, implying that the spotty look in Fig.\ref{fig2}A has
a quantum origin.  As we have done for the density in the real space, we use similar statistics
to quantify the spotty or random image in Fig.\ref{fig2}A.
We first divide the whole momentum space into a series of rings with equal
small spacing and then cut each ring into small pieces of the same area.
In this way, we obtain a set of momentum densities. This set of densities
does not follow the exponential law. However, after renormalizing
the densities of each ring with the averaged density for the same ring,
we again find an exponential distribution as shown in Fig.\ref{fig2}B. \\

In another aspect, following the spirit in Ref.\cite{Tomsovic1993PRE}, we construct
a semiclassical wave function with the classical
momentum distribution $G$%
\begin{equation}
\Psi \left( x,y,t\gg T_{s}\right) \propto\int dv_{x}dv_{y}G\left(
\overrightarrow{v},t\gg T_{s}\right) e^{-i\left( v_{x}x/2+v_{y}y/2+\varphi
_{r}\left( v_{x},v_{y},t\right) \right) }\,,  
\label{randomwavefunction}
\end{equation}%
where $\varphi _{r}\left(v_{x},v_{y},t\right) $'s are assumed to be random phases
caused by the classical chaos.  Since $G$ does not fully describe the quantum
momentum distribution,  the so-constructed wave function $\Psi$ should be
different from the wave function shown in Fig. \ref{fig1}A4. This is indeed the
case: the density computed with the above wave function (Fig.\ref{fig2}C) looks different in
structure from Fig. \ref{fig1}A4 although both of them look random.
However, despite this difference, the random wave function $\Psi$ in Eq.\ref{randomwavefunction} 
also obeys the exponential law (black solid line in Fig.\ref{fig2}D). Due to
the large sampling size, the numerical result fits the exponential function $\exp(-n_0)$
almost perfectly. \\

All the above numerical results point to the universality of the exponential distribution.
This will become apparent as we give a rigorous proof  in the following.
We divide the region of the dynamical evolution into $N$ equal
parts. Then the quantum state $|\Psi \left( t\right) \rangle $ can be
approximated as%
\begin{equation}
\left\vert \Psi \left( t\right) \right\rangle \approx \sum_{j=1}^{N}\alpha
_{j}\left( t\right) \left\vert x_{j},y_{j}\right\rangle .
\end{equation}%
Here $\alpha _{j}(t)=\int_{\Xi _{j}}\Psi \left( x,y,t\right) dxdy/\sqrt{S_{%
\mathrm{total}}/N}$ with $\Xi _{j}$ denoting the $j$th part. Obviously, $%
\alpha _{j}$'s are complex numbers, satisfying the normalization condition%
\begin{equation}
\sum_{j=1}^{N}\left\vert \alpha _{j}\right\vert ^{2}=1\text{.}  \label{norm}
\end{equation}%
We assume that for a fully chaotic  classical system, its corresponding
quantum dynamics will always drive the system to states, where the $\alpha
_{j}$'s are a set of random complex numbers that satisfy the above
normalization condition. With these considerations, the probability of $%
\left\vert \alpha _{j}\right\vert ^{2}$ being between $\gamma _{j}$ and $%
\gamma _{j}+d\gamma _{j}$ is%
\begin{equation}
P\left( \gamma _{j}\right) d\gamma _{j}=\frac{\int d^{2}\alpha _{1}\cdots
d^{2}\alpha _{N}\delta \left( \gamma _{j}-\left\vert \alpha _{j}\right\vert
^{2}\right) \delta \left( 1-\sum_{i=1}^{N}\left\vert \alpha _{i}\right\vert
^{2}\right) }{\int d^{2}\alpha _{1}\cdots d^{2}\alpha _{N}\delta \left(
1-\sum_{i=1}^{N}\left\vert \alpha _{i}\right\vert ^{2}\right) }d\gamma _{j}.
\end{equation}%
It is easy to find that $\int_{0}^{1}P\left( \gamma _{j}\right) d\gamma
_{j}=1$. After straightforward calculations\cite{Ullah1964NPh}, we have%
\begin{equation}
P\left( \gamma _{j}\right) =\left( N-1\right) \left( 1-\gamma _{j}\right)
^{N-2}.
\end{equation}%
In the limit of $N\rightarrow \infty $, we have $\alpha _{j}(t)\approx \Psi
(x_{j},y_{j},t)/\sqrt{n_{s}N}$ and the density probability distribution for
the $j$th part becomes $P\left( n_{j}\right) =e^{-n_{j}}$, where $%
n_{j}=|\Psi (x_{j},y_{j})|^{2}/n_{s}$. Because different part satisfies the
same exponential probability distribution, the dimensionless density
probability distribution $P_{0}\left( n_{0}\right) $ defined for the whole
region satisfies the exponential probability distribution $e^{-n_{0}}$. \\

The above rigorous proof indicates that this exponential law is universal
and should appear in any quantum chaotic systems and many different settings, for example,
in the momentum space. The proof can be used straightforwardly or with small variation
to understand three sets of numerical results presented before the proof.
With the universality, it is not surprising to notice that
this exponential law is known to exist in the speckle pattern of light \cite{ShengpingBook}.
However, it seems that this exponential law has been limited to this special system until now,
and has never been fully discussed in a wide context or explored for broad applications.\\

\begin{figure}[!htp]
\begin{minipage}{5.5cm}
\includegraphics[width=5cm,angle=0]{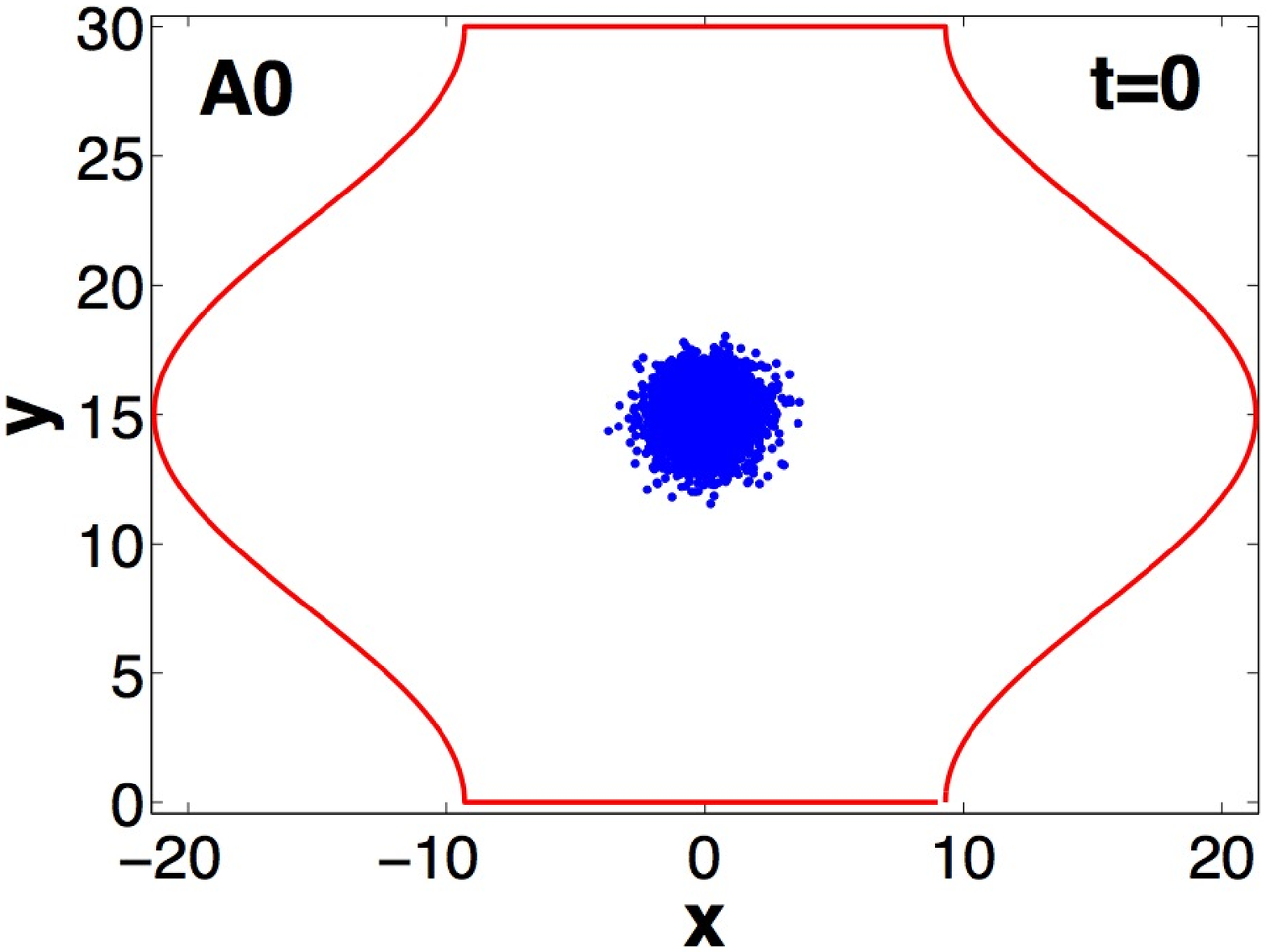} %
\end{minipage}
\begin{minipage}{5.5cm}
\includegraphics[width=5cm,angle=0]{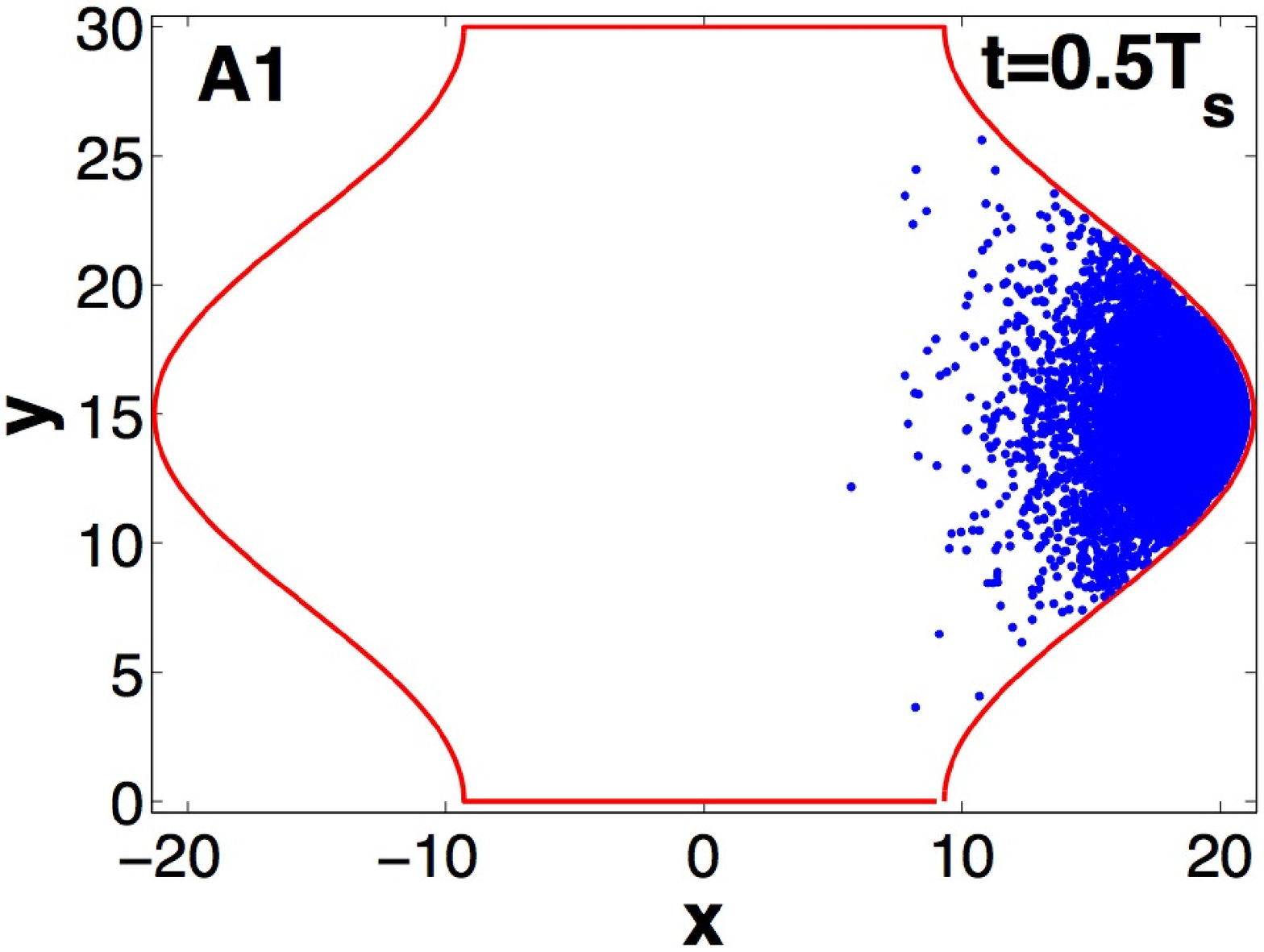} %
\end{minipage}
\begin{minipage}{5.5cm}
\includegraphics[width=5cm,angle=0]{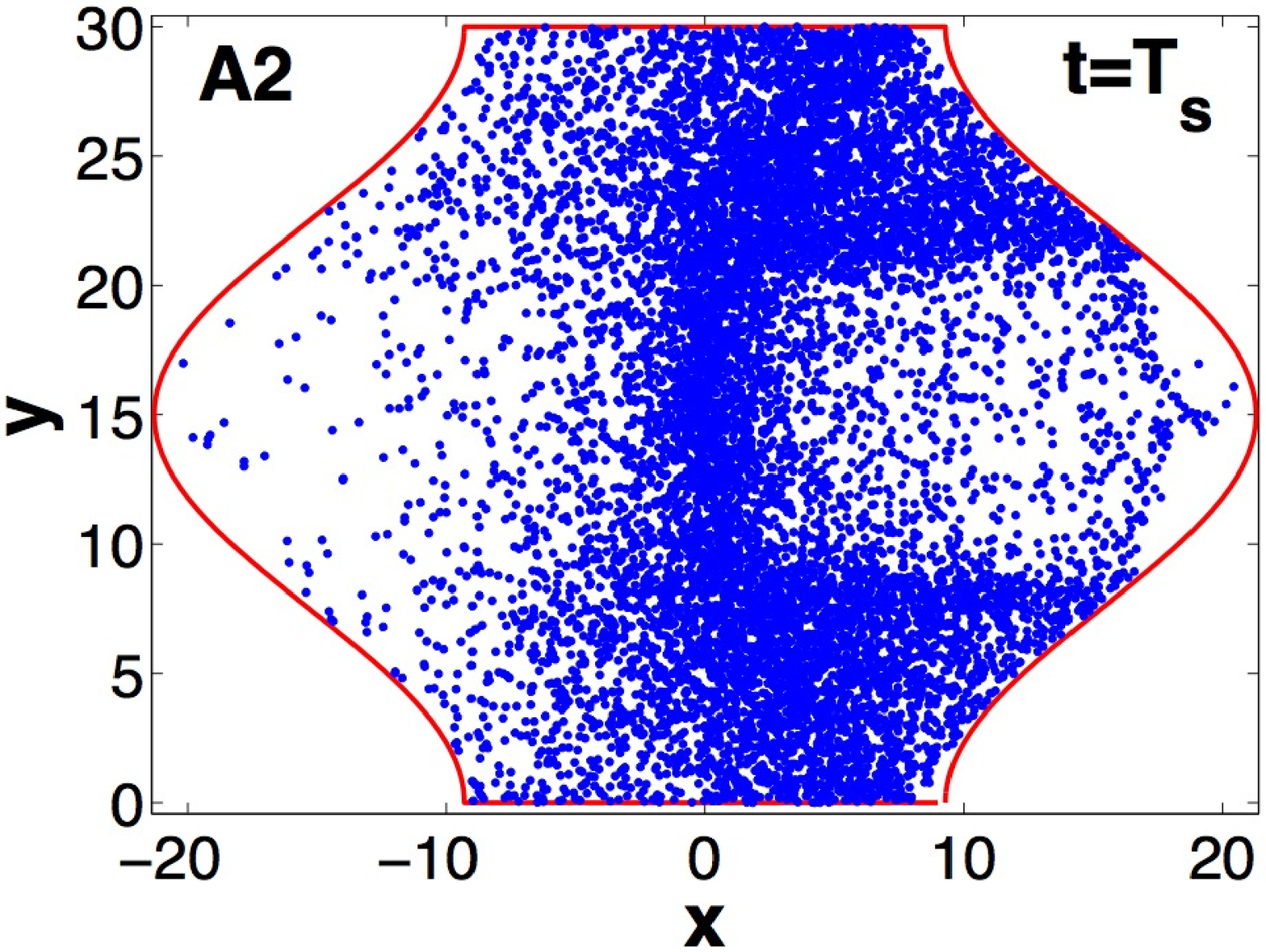}
\end{minipage}
\newline
\begin{minipage}{5.5cm}
\includegraphics[width=5cm,angle=0]{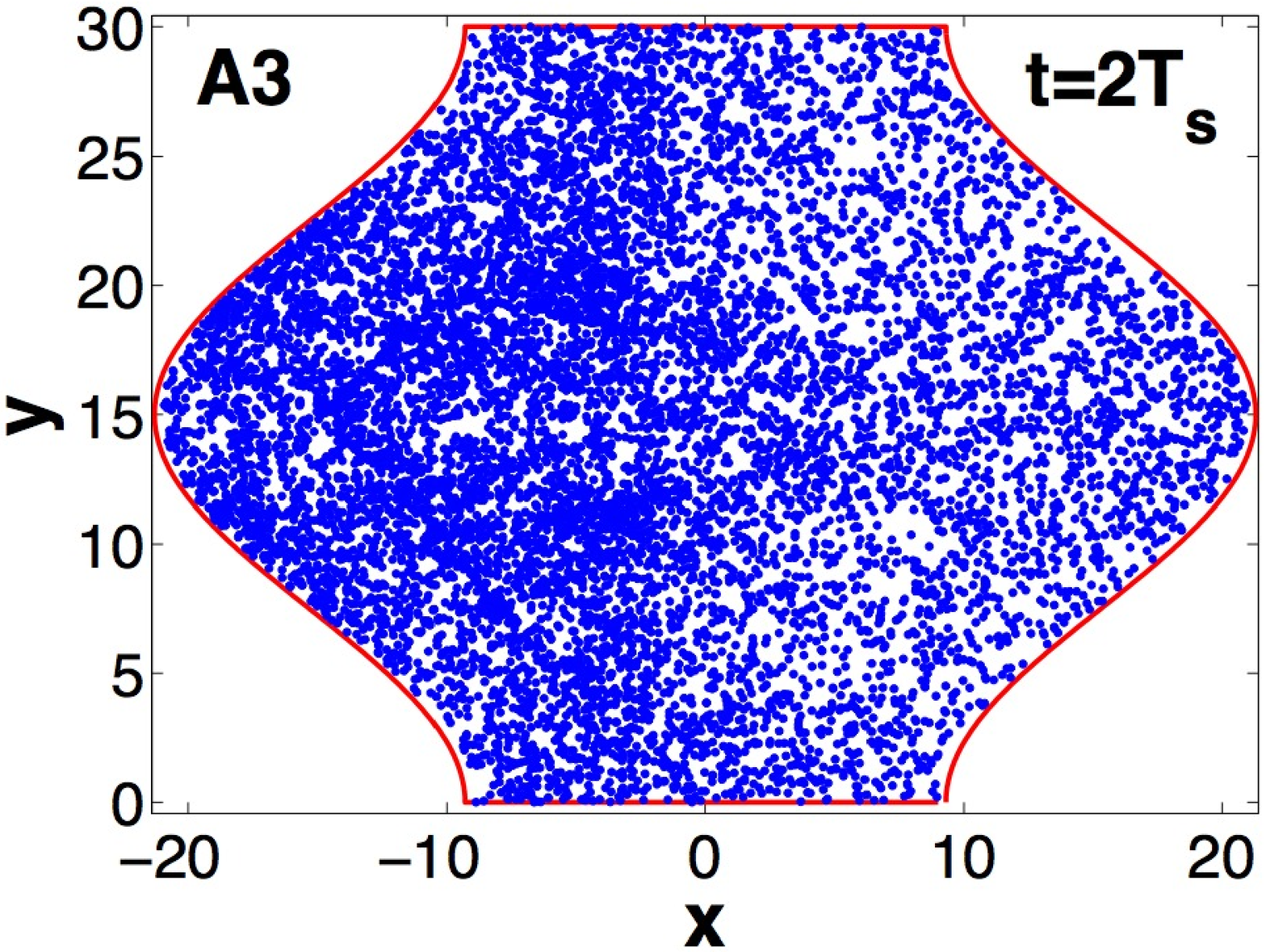} %
\end{minipage}
\begin{minipage}{5.5cm}
\includegraphics[width=5cm,angle=0]{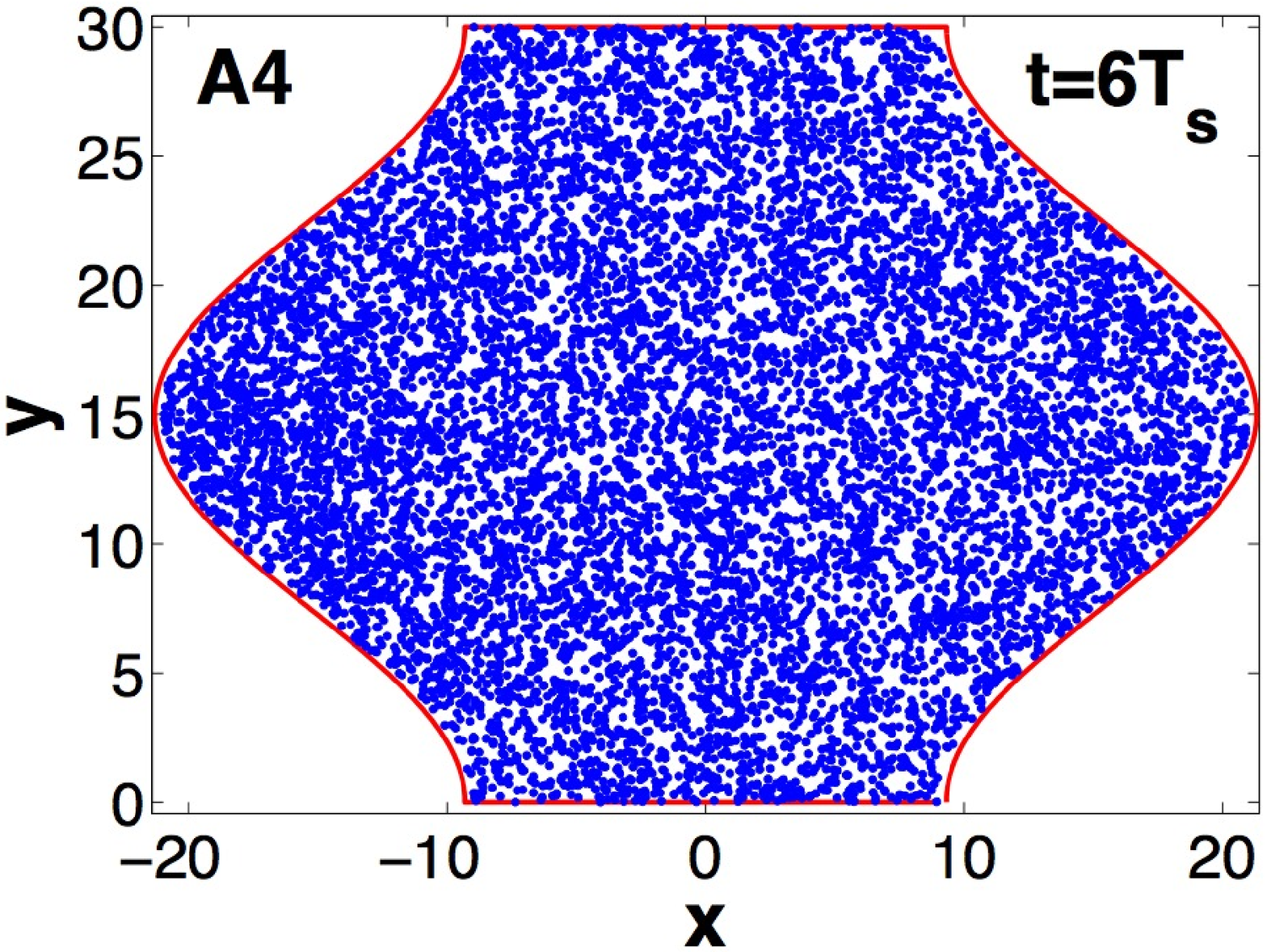} %
\end{minipage}
\begin{minipage}{5.5cm}
\includegraphics[width=4cm,angle=0]{f5}
\end{minipage}
\caption{ Evolution of a cloud of classical particles. (\textbf{A0}%
-\textbf{A4}) are plotted with 20000 particles while the statistics shown in
(\textbf{B}) is done with 1 million particles. The cloud initially has the
exact same spatial and velocity distribution as the initial wave packet in
Fig.\protect\ref{fig1}. The unit cell used for obtaining the statistics  in (\textbf{B}) is 0.6$\times$0.6.}
\label{fig3}
\end{figure}

The established exponential law is in fact a quantum phenomenon. We can
imagine to throw many, many classical particles into the billiard with equal
probability to every part of the whole region. By the central limit theorem,
the resulted density probability distribution is a Gaussian with its peak
located at the averaged density, drastically different from the exponential
distribution for a quantum gas. To confirm this, we have simulated the
evolution of a cloud of non-interacting classical particles with the same spatial and
velocity distributions as the wave packet studied in Fig.\ref{fig1}. The
results are shown in Fig.\ref{fig3}, where we see the classical cloud
expands and reflects very much like its quantum counterpart, the wave packet
in Fig.\ref{fig1}. However, the density probability distribution function
for the classical cloud after long time evolution is Gaussian (see Fig.\ref%
{fig3}B) as we just expected. As a result, we can say that the exponential
distribution is a quantum effect and it can be used in experiment to tell whether a gas is
quantum or classical. \newline


\begin{figure}[!ht]
\includegraphics[width=0.37\linewidth,angle=0]{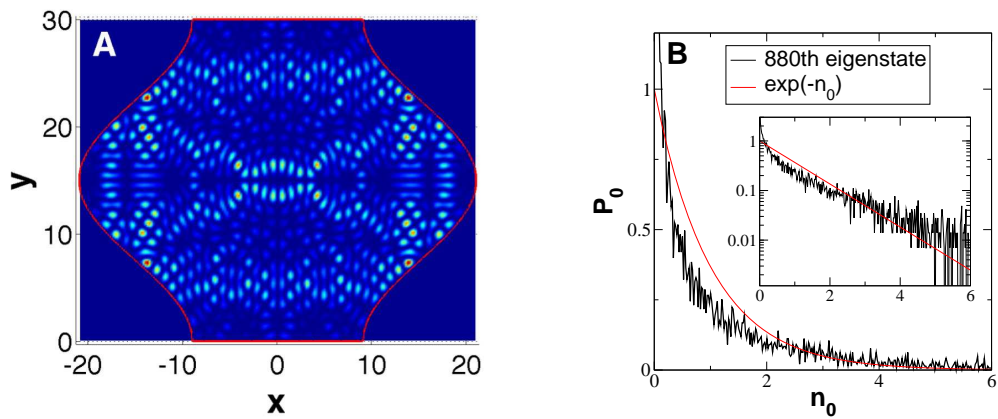} \hspace{1cm} %
\includegraphics[width=0.3\linewidth,angle=0]{pw880.eps}
\caption{The 880th eigenstate of the ripple billiard and its density
probability distribution. The parameters for the ripple billiard are the
same as in Fig.\protect\ref{fig1}. The inset in (\textbf{B}) is the semilog
plot of the same data.}
\label{fig4}
\end{figure}
Eigenstates of a quantum chaotic system have been widely studied. Many of
these eigenstates tend to scar over classical periodic orbits\cite%
{Heller1984PRL}. This means that eigenstates have some regular pattern and
it should not have the exponential distribution. We have checked this
numerically and one example is shown in Fig.\ref{fig4}. In fact, the density
probability distribution of an eigenstate was studied in literature in terms
of amplitude probability distribution and no similar universal behavior was
found\cite{Backer2007EPL}.\newline

The central question in the study of quantum chaos is how the classicality
observed in our daily life emerges from the underlying quantum world
\cite{Gutzwiller1992SciAm,Zurek2001Nature}.
Therefore, it is very interesting to see how the classical Gaussian
distribution emerges from the quantum exponential distribution. We find that
the quantum-classical correspondence in this regard is built by coarse
graining. We have computed the density of the random wave in Eq.(\ref%
{randomwavefunction}) for an area of $400\times 400$. To obtain the density
probability distribution $P_0$, we then divide the whole area equally into
small square cells and sample the averaged densities in these small cells.
We find that the distribution function changes its shape with the sizes of
the cells: as the cell size increases, the distribution changes from
exponential to Gaussian as shown in Fig.\ref{fig2}D. The same transition
is observed for the density in Fig.\ref{fig1}A4 with coarse graining although
the fluctuations are large due to the small sampling size. 
Besides its fundamental
significance, this coarse graining result has also implications in practice:
with a camera of low resolution, one is likely to observe a Gaussian
distribution even for a quantum gas.\\

We have been calling the quantum state reached after long-time evolution
``equilibrium" state. This is because we can not resist the temptation to
compare it with the equilibrium state in thermodynamics. In thermodynamics,
a system will always evolve into an equilibrium state after fully relaxed.
After the equilibrium is reached, the state still changes microscopically;
however, the corresponding macroscopic observables do not change, in
particular, the statistical distribution function no longer changes. This is
the same here for the long-time quantum state: after a certain long time ($%
\sim 6T_s$ in our case), the quantum state still changes locally while its
overall structure remains the same and the density probability distribution function 
sticks to the exponential
law. It is worthwhile to compare the hyper-sphere in the Hilbert space
defined by the normalization condition (\ref{norm}) to the hyper-sphere in
the phase space defined by a constant energy. As we know, for almost all the
microscopic states in the phase space sphere, they share the same
macroscopic characteristics, such as the same pressure and the same thermal
expansion coefficient. Similarly, with the understanding that we have gained
so far, almost all the states in the Hilbert space should have the
exponential distribution. It is true that there are many states such as the
eigenstate shown in Fig.\ref{fig4}, which do not have the exponential
distribution. However, we expect the measure of these states is zero. In light of
this discussion, it is likely that this exponential distribution is related to the
thermalization of isolated quantum systems \cite{Rigol2008Nature}. \newline

Quantum chaotic system has been defined as the quantum system whose
corresponding classical system is chaotic. With the exponential
distribution law, we may be now ready to define quantum chaotic
system without referring to classical system. One might choose to
define a quantum chaotic system as a quantum system which has the
ability to drive an initially regular wave packet to a coherent
superposition of completely random wave functions, which has the
observable exponential density probability distribution. We finally
note that for the exponential distribution, the density fluctuation
is $\delta n^{2}=n^{2}$, which is much larger than the Gaussian
distribution. In other words, the quantum fluctuation in density is
much larger than the classical fluctuation. This large quantum
fluctuation in density may help to explain the formation of stars if
a wave function for the whole universe is
assumed\cite{Young2010SAm,Smith2004Book}.

\section*{ACKNOWLEDGMENTS}
This work was supported by National Science Foundation of China under Grant Nos. 10875165,
10825417, 10634060, and NBRPC 2006CB921406.







\end{document}